\documentclass[aps,pre,twocolumn,showpacs,superscriptaddress]{revtex4-1}  % for review and submission
\usepackage{graphicx}  % needed for figures
\usepackage{dcolumn}   % needed for some tables
\usepackage{bm}        % for math
\usepackage{amssymb}   % for math
\usepackage{times}
\usepackage{amsmath}
\usepackage{amsfonts}
\usepackage{amsthm}
\usepackage[normalem]{ulem}

\usepackage{tabularx}
\usepackage{rotating}
\usepackage{multirow}
\usepackage{array}

\usepackage{fancyhdr}
\usepackage{graphicx}
\usepackage{epsfig}
\usepackage[dvipsnames]{xcolor}
\usepackage{tikz}
\usetikzlibrary{arrows,positioning,calc}
\usetikzlibrary{matrix}
\usetikzlibrary{shapes}
\usetikzlibrary{chains}

\usepackage{lipsum}
\usepackage[english]{babel}
\usepackage[utf8]{inputenc}
\usepackage[T1]{fontenc}
\usepackage{lmodern}
\usepackage{subfigure}
\usepackage{titlesec}
\usepackage{mathtools} %xxx
\usepackage{tcolorbox} %xxx
\usepackage{url}
\usepackage{epstopdf}
\usepackage{bigints}
\newcommand{\mathd}{\mathrm{d}}
\newcommand{\ovl}[1]{\overline{#1}}
\newcommand{\til}[1]{\widetilde{#1}}
\newcommand\ddfrac[2]{\frac{\displaystyle #1}{\displaystyle #2}}
\DeclareMathOperator\arctanh{Arctanh}
\begin{document}

\widetext
% the following line is for submission, including submission to the arXiv!!
%\hspace{5.2in} \mbox{Fermilab-Pub-04/xxx-E}

\title{Continuous nonequilibrium transition  driven by the heat flow}

\author{Yirui Zhang }
\affiliation{Institute of Physical Chemistry, Polish Academy of Sciences, Kasprzaka 44/52, PL-01-224 Warsaw, Poland}

\author{Marek Litniewski }
\affiliation{Institute of Physical Chemistry, Polish Academy of Sciences, Kasprzaka 44/52, PL-01-224 Warsaw, Poland}

\author{Karol Makuch}
\affiliation{Institute of Physical Chemistry, Polish Academy of Sciences, Kasprzaka 44/52, PL-01-224 Warsaw, Poland}

\author{Pawel J. {\.Z}uk}
\affiliation{Institute of Physical Chemistry, Polish Academy of Sciences, Kasprzaka 44/52, PL-01-224 Warsaw, Poland}
\affiliation{Department of Physics, Lancaster University, Lancaster LA1 4YB, United Kingdom}

\author{Anna Macio\l ek }\email{maciolek@is.mpg.de }
\affiliation{Institute of Physical Chemistry, Polish Academy of Sciences, Kasprzaka 44/52, PL-01-224 Warsaw, Poland}
\affiliation{Max-Planck-Institut f{\"u}r Intelligente Systeme Stuttgart, Heisenbergstr.~3, D-70569 Stuttgart, Germany}

\author{Robert Ho\l yst}\email{rholyst@ichf.edu.pl }
\affiliation{Institute of Physical Chemistry, Polish Academy of Sciences, Kasprzaka 44/52, PL-01-224 Warsaw, Poland}

\date{July 11, 2021}

\begin{abstract}
We discovered an out-of-equilibrium transition in the ideal gas between two walls, divided by an inner, adiabatic, movable wall. The system is driven out-of-equilibrium by supplying energy directly into the volume of the gas. At critical heat flux, we have found a continuous transition to the state with a low-density, hot gas on one side of the movable wall and  a dense, cold gas on the other side. Molecular dynamic simulations of the soft-sphere fluid confirm the existence of the transition in the interacting system. We introduce a stationary state Helmholtz-like function whose minimum determines the stable positions of the internal wall. This transition can be used as a paradigm of transitions in stationary states and the Helmholtz-like function as a paradigm of the thermodynamic description of these states.   
\end{abstract}

\maketitle

Nonequilibrium thermodynamics \cite{groot1962, prigogine, oono1998steady,Sasa:2006,revUdo} have never reached the same status as equilibrium thermodynamics \cite{holyst2012thermodynamics}.  Despite many decades of study, the question concerning the existence of universal extremal principles that determine behaviour of  nonequilibrium systems is still open. The most prominent propositions are the maximum/minimum entropy production principles \cite{onsager1,onsager2,ziegler}. 
There are attempts to provide theoretical justifications of the maximum entropy production principle   based on information theory \cite{PhysRevE.80.021113,Dewar2014} or least action principle \cite{wang2006maximum}. However, applicability of both principles is widely discussed \cite{martyushev2013entropy,dewar2014beyond,Dewar2014,endres2017entropy,doi:10.1063/1.2400859}  and the lack of their predictive success is acknowledged.
On the other hand, there is  a significant progress in the characterization of nonequilibrium systems  by fluctuation theorem that involve probability distribution of quantities defined on trajectories~\cite{revUdo, revUdo01, gallavotti1995dynamical, lebowitz1999gallavotti, kurchan1998fluctuation, evans1994equilibrium, jarzynski, crooks2000path}. Although stochastic dynamics links the statistics of trajectories with the entropy production, to our knowledge, this approach has not been used for predictions of steady states.

Equilibrium thermodynamics provides a clear definition of a few macroscopic variables defining the equilibrium state and function, which has a minimum at this state. For example, the state of a one-component system interacting with the environment via isothermal walls is defined by three parameters $T$ temperature, $V$ volume, and $N$ number of particles. The state's function, the Helmholtz free energy, $F(T, V, N)$, is minimized in the equilibrium state. The minimization is over potential states obtained at constant $T, V, N$ via internal constraints. 
 The present  paper introduces a methodology of nonequilibrium thermodynamics having
a similar structure as the equilibrium counterpart. We use this methodology to analyze the continuous transition between two nonequilibrium stationary states that we discovered in a paradigmatic heat flow model. 

In a series of our recent papers \cite{Robert, zhang2020energy, zhang2020storage}, we have analyzed one-component systems subjected to the constant heat flow. In the system's stationary state, its internal energy is a function of $T, L, N$, and the heat flux, $J$. Here $T$ is the temperature at the boundary, where the heat flux leaves the system; $L$ is the size of the system, and $N$ is the number of particles.
This observation suggests that the thermodynamic parameters describing the stationary state of such a nonequilibrium steady state are similar to those describing its equilibrium counterpart.  A new thermodynamic parameter characterizing the state is the heat flux, $J$.

\begin{figure}[htb!]
\includegraphics[width=0.4\textwidth]{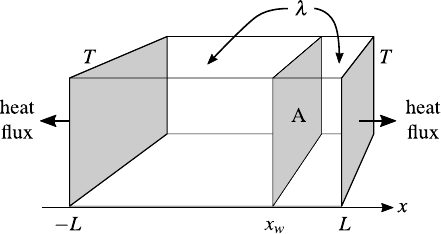}
\caption{Schematic plot of a system with a movable wall. The vertical black lines represent the wall. The left and right boundaries have an area of $\mathrm{A}$ and are placed at $\pm L$. The heat flows through the boundaries. Energy is supplied through an homogeneous external energy input of density $\lambda$. The external walls are kept at temperature $T$.}
\label{fig:1}
\end{figure}

Here, we study an  ideal gas between two parallel walls  at fixed temperature $T$ separated by the distance $L$. The energy flows into the system's volume in the form of  heat, and  the energy supplied into the system per unit time and unit volume is $\lambda = J/V$. Such an energy supply can be realized by microwaves in an appropriate designed  experimental setup. 
A schematic plot of the system is shown in Fig.~\ref{fig:1}. The internal energy  in the steady state $U$ has the following form: 
\begin{equation}
U = U_{eq} f(\lambda L^{2}/kT), \label{eqn:tot-e}
\end{equation}
where  $k$ is the thermal conductivity and $U_{eq}$ is the energy of the same system in the absence of external energy supply. We introduce a movable adiabatic wall parallel to the bounding walls. At equilibrium, the wall is located precisely in the middle of the system. For small heat fluxes, the position of the wall is stable. Above a critical flux, the wall moves towards one of the bounding surfaces.  We show that the minimum of the nonequilibrium Helmholtz-like free energy, defined in this paper, determines the stationary state's wall position. 
Unlike in the existing  approaches, see  e.g. Refs~\cite{Niven2010,komatsu2008steady}, our construction of a nonequilibrium free-energy-like potential does not rely on the knowledge of  entropy.

As shown in Fig.~\ref{fig:1}, the left and right boundaries are  fixed at $x=\pm L$, with a large area $\mathrm{A} \rightarrow \infty$, giving $V = 2\mathrm{A}L$. 
A movable wall is adiabatic, i.e.,  does not allow heat to pass it,  and constitutes the internal constraint. Thus the system is separated into two subsystems $1$ and $2$, each with a fixed number of particles $N_{1}$ and $N_{2}$. In the following, we will denote variables of the subsystem $i = 1,2$ with subscript $-_{i}$, and the subsystem on the left(right) hand side is designated $1(2)$.
The wall is assumed to move freely without friction. Therefore, the condition for the total system to reach steady state is that the pressures exerted by each subsystems are equal $P_{1}(x_{w}) = P_{2}(x_{w})$, where $x_{w}$ is the position of the wall. 
In equilibrium, according to the ideal gas law 
$P_{eq}V = N k_{B}T$,  
where $k_B$ is the Boltzmann constant, therefore the intended ratio of $N_{1}/N_{2}$ is achieved by placing wall such that $N_{1}/N_{2}$ is  equal  to the ratio of the volumes $V_{i}$. 
We make two assumptions about this system. 
First, we assume that the heat conduction follows Fourier's law. Then, the temperature profile can be obtained from the local continuity equation of energy 
\begin{equation} 
- k \nabla^{2}T(\vec{r}) = \lambda,   
\label{eqn:laplace}
\end{equation}
Second, assuming local equilibrium and the equation of state for monoatomic ideal gas  can be extended to be valid locally so that  in nonequilibrium steady states 
%\begin{align}
\begin{equation}
\label{eqn:ss-state-eqn}
 P = n(\mathbf{r}) k_{B} T(\mathbf{r}) \qquad
\epsilon = \dfrac{3}{2} n(\mathbf{r})k_{B}T(\mathbf{r}) = \dfrac{U}{V} ,  
%\end{align}
\end{equation}
where $n(\mathbf{r})$ is the local particle number density at position $\mathbf{r}$, with $\int_{V} n(\mathbf{r}) \mathd^3 r = N$; $T(\mathbf{r})$ is the local temperature at $\mathbf{r}$; $\epsilon$ is the energy density. In the above two relations, both $P$ and $\epsilon$ are constant across the system.
This is because the redistribution of mechanical energy occurs much faster than the redistribution of heat.
From the above two assumptions, the steady state energy density of the system  and of each subsystem can be obtained using 
\begin{equation}
 \epsilon =  \dfrac{3}{2}N k_{B}\ddfrac{1}{\int_{V} \ddfrac{\mathd \vec{r}}{T(\vec{r})}} \qquad
\epsilon_{i} =  \dfrac{3}{2}N_{i} k_{B}\ddfrac{1}{\int_{V_{i}} \ddfrac{\mathd \vec{r}}{T_{i}(\vec{r})}},
\label{eqn:e}  
\end{equation}
where the temperature profiles are obtained from Eq.~(\ref{eqn:laplace}) with the appropriate boundary conditions. As a result, the energy of the system prior to the constraint is given by Eq.~(\ref{eqn:tot-e})
and the energy of the subsystem of $N_{i}$ particles under the constraint is 
\begin{equation}
U_{i} = U_{i,eq} f(\lambda L_{i}^{2}/kT) = \dfrac{3}{2}N_{i}k_{B}T f(\lambda L_{i}^{2}/kT) 
\label{eqn:sub-e}
\end{equation}
with $N_1+N_2 = N$, where $U_{eq}=(3/2)Nk_{B}T$ and $U_{i, eq}$ are the system and subsystem energy in equilibrium, $L_{i}$ is the length of the subsystem with $L_{1} = L+x_{w}$ and $L_{2} = L-x_{w}$, and the function $f$ is given by  
\begin{equation}
f(x) \equiv \ddfrac{\sqrt{x(x+2)}}{2\arctanh \sqrt{x/(x+2)}}. 
\end{equation}
The derivations are shown in the Appendix~\ref{appen1}. 
We would like to make two remarks. First, the variables of $f$ is separated to $\lambda$ and $L_{i}^2/kT$, where $\lambda$ is the control parameter, and the coefficient $L^2/kT$ (or $L_{i}^2/kT$ for the subsystems) are parameters that are either of the intrinsic properties of the system, or of the environment that is not changed ($T$).  Second, this model is seemingly similar to the model considered  in our previous paper \cite{Robert} (named  there as \textit{case 1}). In \textit{case 1}, however, the adiabatic wall is fixed in space, and the subsystems are independent. Whereas in the movable wall model, the constraint couples the two subsystems. This single difference results in an interesting second-order nonequilibrium phase transition which we will discuss next. 

\begin{figure}[htb!]
\includegraphics[width=0.46\textwidth]{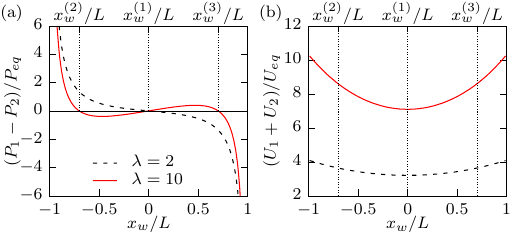}
\caption[option]{System's response as a function of $x_w$ for two values of $\lambda$.
 (a) The difference between pressures in compartments normalized with equilibrium pressure $P_{eq}=Nk_bT/V$.
 (b) Total internal energy of the system normalized with $U_{eq}$.
 The vertical lines mark the position of the steady states $x^{(1)}_w$ for $\lambda=2$ and $x^{(1)}_w,x^{(2)}_w$ and $x^{(3)}_w$ for $\lambda=10$.
$\lambda$ is in units of $kT/L^2$}
\label{fig:mov}
\end{figure}

For our movable wall model, the condition of nonequilibrium steady states can be equivalently stated as
$P_{1}(x_{w}) = P_{2}(x_{w})$ or 
$\epsilon_{1}(x_{w}) = \epsilon_{2}(x_{w})$. The solutions $x_{w}^{(i)}$ (where the superscript $^{(i)}$ indicates the $i$th solution) are obtained numerically. Graphically, the solutions and their corresponding stability can be observed at and around the   zeros of $P_{1}(x_{w}) - P_{2}(x_{w})$. 
We set $N_{1} = N_{2} = N/2$ and observe that as $\lambda$ increases (at fixed $T$, $ V$ and $N$), the system undergoes a second-order nonequilibrium phase transition. The order parameter is the stable position of the wall $x_{w}$. 
For small $\lambda$, we find a stable steady state at $x_{w}=0$. As this division gives identical subsystems, $P_{1}=P_{2}$ trivially. An example  is shown in Fig.~\ref{fig:mov}(a) ( dashed curve). 
One can see that the curve $P_{1}(x_{w}) -P_{2}(x_{w})$ is monotonic and exhibit a single zero-crossing point at $x_{w}=0$. To evaluate the stability, suppose now that the constraint is pushed away from the center towards $x_{w} > 0$. One observes that $P_{1} - P_{2} < 0$. Consequently, the pressure difference will push the wall back towards $x_{w}^{(1)} = 0$. Therefore, $x_{w}^{(1)} = 0$ is a stable solution in this phase. 
For large $\lambda$, interestingly, the system has three  steady states. They  correspond to the position of the wall at  $x_{w}^{(1)}=0$ and at  $x_{w}^{(2)}=-x_{w}^{(3)} \neq 0$ due to symmetry. Qualitatively, one can imagine the asymmetric case where, according to Eq.~(\ref{eqn:ss-state-eqn}), the smaller average number of particle density $\ovl{n}_{i}\equiv N_{i}/V_{i}$ for the  larger subsystem is compensated with a higher overall temperature, whereas larger $\ovl{n}_{i}$ in the smaller system is compensated with a lower overall temperature. This compensation can occur because the heat flux of the movable wall model is proportional to the volume $V_{i}$ (inverse proportional to $\ovl{n}_{i}$). An example of three solutions is shown in Fig.~\ref{fig:mov}(a) (zeros of the red curve). 
Among these $3$ solutions, further analysis show that $x_{w}^{(2)}$ (and $x_{w}^{(3)}$) are stable, whereas $x_{w}^{(1)}$ is unstable. As we can see from Fig.~\ref{fig:mov}(a), the  red curve is no longer monotonic and exhibits zero-crossing at $3$ points. Suppose now that the constraint is pushed away from the steady state, in one case to the position between $x_{w}^{(1)}$ and $x_{w}^{(2)}$, and in the other case beyond $x_{w}^{(2)}$. In both situations, the pressure difference will push the constraint towards $x_{w}^{(2)}$. Due to symmetry, the same argument holds for $x_{w}^{(3)}$ when the starting point of the constraint is $x_{w}>0$.  Note that the total energy of the system $U_{1}+U_{2}$ has minimum always at $x_{w}=0$ - see Fig.~\ref{fig:mov}(b).

\begin{figure}[tb!]
\includegraphics[width=0.5\textwidth]{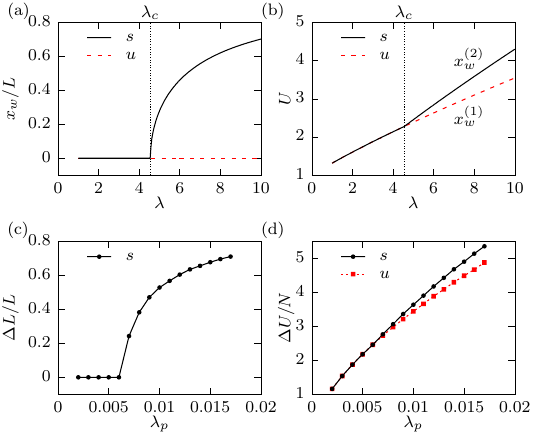}
\caption[option]{(a) Diagram showing the location  of the stationary states in the parameter space  $(x_{w}, \lambda)$. Stable (s)  and unstable (u) stationary states are indicated in black (red); (b) Energy of the total system $U$ (in units of $U_{eq}$) as a function of the flux per unit volume $\lambda$ (in units of $kT/L^2$). $(c)$ and $(d)$ show molecular dynamic simulations results for the soft-sphere fluid: 
$(c)$ the relative shift of the wall $\Delta L/L = \vert \langle x_{w} \rangle/L\vert$ as a function of $\lambda_p$ - the mean rate of energy added per particle. $\langle x_{w}\rangle$ is the mean value of $x_{w}$ at the end of the simulation run. 
(d) $\Delta U/N$ - the deviation of energy per particle (in units of the amplitude $\varepsilon$ of the interaction potential) from its initial value established before the shift of the wall (black circles) and after the shift (red squares) as a function $\lambda_p$. }
\label{fig:mov_U}
\end{figure}

\begin{figure}[tb!]
\includegraphics[width=0.32\textwidth]{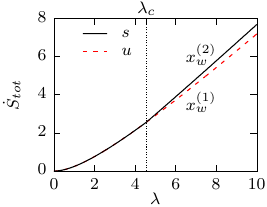}
\caption[option]{ Total entropy production rate $\dot{S}_{tot}$ (in units of $k V/L^2$) as a function of the flux per unit volume $\lambda$ (in units of $kT/L^2$).   Stable (s)  and unstable (u) stationary states are indicated in black (red). }
\label{fig:mov_EP}
\end{figure}

\begin{figure}[tb!]
\includegraphics[width=0.5\textwidth]{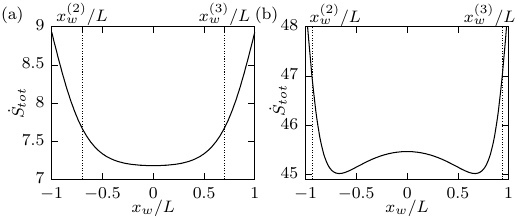}
\caption[option]{ Total entropy production rate $\dot{S}_{tot}$ (in units of $k V/L^2$)   for (a)  $\lambda= 10$  and (b) $\lambda= 50$.
 The vertical lines mark the position of the steady states, which differ from the positions of the extremes of  $\dot{S}_{tot}$.
$\lambda$ is in units of $kT/L^2$}
\label{fig:mov_EP1}
\end{figure}

\begin{figure}[htb!]
\includegraphics[width=0.5\textwidth]{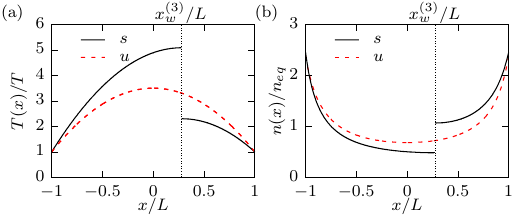}
\caption[option]{ (a) Temperature $T(x)$ (in units of $T$)  and  (b)  number density $n(x)$ (in units of $n_{eq} = N/V$ profiles for $ \lambda = 5.0$, slightly
above the transition point. The stable position of the wall $-x_{w}^{(2)} = x_{w}^{(3)} = 0.278564$ is marked  by vertical lines.
 Stable (s)  and unstable (u) stationary states are indicated in black (red). }
\label{fig:profiles}
\end{figure}

Thw diagram showing the position of stationary states in the parameter space  $x_{w} - \lambda$ space (at fixed $T, V, N$) is presented in Fig.~\ref{fig:mov_U} (a). One can see that upon increasing $\lambda$ the  transition from one steady state to another is continuous. The transition point occurs at $\lambda_{c} L^{2}/kT \approx 4.55344$ (see Appendix~\ref{app:critical-point}). 
The steady state energy of the total system $U = U_{1} + U_{2}$  is plotted in Fig.~\ref{fig:mov_U}(b). Interestingly, the energy of the stable steady states is higher than the unstable steady state. 
The same is valid for the total entropy production rate $\dot{S}_{tot} = \mathrm{A}\int_{-L}^{L} \sigma_s (x) dx$, where  $\sigma_s= k\left(\partial T(x)/\partial x\right)^2/(T(x))^2$ - see  Fig.~\ref{fig:mov_EP}. 
In Fig.~\ref{fig:mov_EP1} we demonstrate that the transition cannot be predicted from the  extremum of  the total entropy production rate as a function of $x_{w}$.  $\dot{S}_{tot}$ has a single minimum  at $x_{w}=0$ for the values of $\lambda$  above the transition. The two minima occur  at some $\lambda^*$,  which is much  larger  than $\lambda_{c}$,  but the positions of these minima do not correspond to the stable positions of a movable wall.
Above $\lambda_{c}$, the  temperature and number density profiles develop discontinuity at the movable wall as shown in Fig.~\ref{fig:profiles}. 

This phase transition diagram is obtained based on the assumptions that may fail far from equilibrium. In order to test our analytical results
we performed molecular dynamics (MD) simulations \cite{S1-allen1989computer} of the soft-sphere fluid
where no assumptions concerning local equilibrium or constancy of heat conductivity are made.
MD simulations provided qualitatively the same results for the energy storage as a function of the mean rate of energy added per particle $\lambda_p$ and the phase transition is retrieved (see Fig.\ref{fig:mov_U}(c) and (d)).
Simulations are performed for fixed $N = 153600$ particles enclosed in the rectangular box of a size $L_{z} = L_{y} = 275.8\sigma, L_{x} \equiv 2L =  658.3\sigma$, where the molecular size unit  $\sigma$ is set to 1, with periodic boundary conditions  applied  along $y$ and $z$ axis. 
The energy flux is proportional to the  density  \textit{i.e.}, the same amount of energy is added to the same volume and equally shared between all particles in that volume (for more details of simulations see Appendix~\ref{app:simulation}). 

For a system with a volume V and a fixed number of  particles N  in contact with a  heat bath at temperature $T$ and  driven out-of-equilibrium by external control parameter $\lambda$, we propose a nonequilibrium state function $B$ that is minimized for stable steady states, and provide its expression. We will demonstrate the use of this potential using the movable wall model, and show that it predicts the correct stable steady states.  
The development of the nonequilibrium state function  is based on the assumption that the relevant parameters are the thermodynamic variables $T, V, N$, and the parameter $\lambda$  that accounts for the nonequilibrium. 
In the limiting case  $\lambda \to  0$, this state function should agree with the equilibrium free energy 
%\begin{equation}
$\lim_{\lambda \rightarrow 0} B(T, V, N, \lambda) = F_{eq}(T, V, N)$, 
%\label{eqn:post1}
%\end{equation}
which is the correct state function of an  equilibrium system. 
Moreover, in analogy to the equilibrium free energy, we postulate that 
 $B$ satisfies 
\begin{equation}
\mathd B = - S \mathd T - P \mathd V + \mu \mathd N - X \mathd \lambda
\label{eqn:B-diff}
\end{equation}
where $P(T, V, N, \lambda), \quad S(T, V, N, \lambda), \quad \mu(T, V, N, \lambda)$ and $X(T, V, N, \lambda)$ are state functions conjugate to $V, T, N$ and $\lambda$, respectively. 
Equation (\ref{eqn:B-diff}) defines the steady state pressure $P$, the steady state entropy $S$ and the steady state chemical potential $\mu$, which  should retrieve its equilibrium values as $\lambda \rightarrow 0$; $X$ is the new variable purely due to nonequilibrium, which has no equilibrium counterpart. Note that the first three terms are  analogous to the differential form of the equilibrium free energy $dF_{eq} = - S_{eq} \mathd T - P_{eq} \mathd V + \mu_{eq} \mathd N$. 
Finally, we postulate that $X$ is of the form of
\begin{equation}
X \propto  \left( \ddfrac{U - U_{eq}}{\lambda} \right). 
\label{eqn:x}
\end{equation}
This is inspired by our  earlier observations \cite{Robert, zhang2020storage, zhang2020energy} that for several seemingly different systems, a quantity $\mathcal{T} = (U - U_{eq})/J_{U}$ is minimized for steady states, where $J_{U}$ is the total heat flow. The two quantities $X$ and $\mathcal{T}$ are similar since $\lambda$ is quantitatively the total heat flow per unit volume. The proportionality constant  is obtained through an argument of consistency that  we describe below. 
Now, we proceed to demonstrate the use of $B$ in the movable wall model. Taking the energy expression (\ref{eqn:tot-e}),  we have used consistency relations analogous  to the Maxwell relations of equilibrium thermodynamics 
in order to obtain the  expression for the nonequilibrium potential and the
steady state functions of the unconstrained system (see Appendix~\ref{app:consist-check}); the proportionality constant in Eq.~(\ref{eqn:x}) equal to $1/3$ restores the pressure correctly. We found:
\begin{equation}
B(T, V, N, \lambda) = F_{eq}(T, V, N) - \int_{0}^{\lambda} X(T ,V, N, \lambda') \mathd \lambda'. 
\label{eqn:B}
\end{equation} 
\begin{equation}
\begin{aligned}
S & = \ddfrac{N k_{B}}{2}\int_{0}^{\lambda} \left(f(\lambda' L^{2}/kT) - 1\right) \ddfrac{\mathd \lambda'}{\lambda'} \\
&  - \ddfrac{N k_{B}}{2} (f (\lambda L^{2}/kT) - 1) + S_{eq} \\
\end{aligned}
\label{eqn:ss-s}
\end{equation}
\begin{align}
P &= \ddfrac{N k_{B} T f(\lambda (L^{2}/kT))}{V} 
				       = \ddfrac{3}{2} \ddfrac{ U}{V} \label{eqn:ss-p}\\
\mu &= \ddfrac{k_{B}T}{2}\int_{0}^{\lambda} \left(f(\lambda' (L^{2}/kT)) - 1\right)\ddfrac{\mathd \lambda'}{\lambda'}  + \mu_{eq} \label{eqn:ss-mu}
\end{align}
Note that as $\lambda \rightarrow 0$, $f \rightarrow 1$. 
It is then obvious that from these four expressions we retrieve the correct corresponding equilibrium potentials in the limit of  $\lambda \rightarrow 0$.
From  Eqs.~(\ref{eqn:ss-s}), (\ref{eqn:ss-p}) and (\ref{eqn:ss-mu}), we also obtain the integral form of $B$ of the unconstrained system as 
\begin{equation}
B = U - TS - 4X\lambda,  
\label{eqn:b-int}
\end{equation}
in consistency with Eq.~(\ref{eqn:B}). 
This form is again analogous to the equilibrium free energy expression $F_{eq} = U_{eq} - TS_{eq}$. The additional term is the conjugate pair due to nonequilibrium $X\lambda$ with a coefficient $4$. 

For the constrained system, 
the nonequilibrium potential of  movable wall system is given by
\begin{align}
B(T, V, N_{1}, N_{2}, x_{w}, \lambda)& = F_{eq}(T, V, N_{1}, N_{2}, x_{w}) \\
&- \int_{0}^{\lambda} X(T, V, N_{1}, N_{2}, x_{w},\lambda') d\lambda', 
\end{align}
where 
\begin{equation}
X(T, V, N_{1}, N_{2}, x_{w}, \lambda) = \dfrac{1}{3} \ddfrac{U_{1}+ U_{2} - U_{eq}}{\lambda}. 
\end{equation}
The extremum condition at fixed $T, V, N_{1}, N_{2}$ and  $\lambda$ reduces  to the condition of matching pressure:
\begin{equation}
\dfrac{\partial{B}}{\partial{x_{w}}}\bigg|_{x_{w}^{*}} = 0 \Leftrightarrow - \mathrm{A} (P_{1} - P_{2}) = 0, 
\end{equation}
where 
\begin{align}
&P_{1} = \dfrac{N_{1}k_{B}T }{V_{1}} f(\lambda \dfrac{(L + x_{w})^2}{kT}), \\
&P_{2} = \dfrac{N_{2}k_{B}T }{V_{2}} f(\lambda \dfrac{(L - x_{w})^2}{kT}). 
\end{align}
Thus, we have demonstrated that the extremum points correctly predict the steady states in the movable wall model. Further, analysis shows that $x_{w}^{*} = 0$ corresponds to a local maximum and $x_{w}^{*} \neq 0$ local minimum (see Appendix~\ref{app:extremum}).

In conclusion, for the movable wall case, we have retrieved the $3$ steady states as the extremum of $B$, and the minimum of $B$ predict correctly the stable steady state. 
In a general case of  $N_{1} \neq N_{2}$ the system still exhibits a second order phase transition under certain circumstances.
The behaviour is more complex as it involves an additional variable and needs further study.  
The movable wall model studied here analytically exhibits  second-order nonequilibrium phase transition. 
The transitions in out-of-equilibrium states that can be fully characterized by analytical calculations are extremely rare.
Therefore the  transition that we have found can be used as a paradigm of such transitions. We have provided 
a full thermodynamic description of the transition introducing the Helmholtz-like function for stationary states.
We think that such a description analogous to ordinary thermodynamics has great potential in the description of  stationary states and could push forward nonequilibrium thermodynamics.

Concerning the  physical realization of the volumetric energy supply, the following example of ''gedanken`` experiment shows that in principle it is possible to deliver the same amount of energy per unit volume into the system.
Such delivery requires microwave device.
The total flux of photons is  $I_V$ (each of energy $e$), and it enters a given volume $V$. $I_V$ is adjustable and the microwave device is constructed in such a way as to deliver a predetermined flux to the chosen sub-volume of our system.
Some external device measures (e.g. by fluorescence) the number of molecules, $N_V$, in a given volume $V$.
The external device is coupled to the microwave device. This coupling allows to change the flux at will, depending on the number of molecule $N_V$.
Each molecule has a fixed probability to capture one photon given by $p$.
Now $\lambda=I_V\times N_V \times p \times e/V=const$. For a fixed $V$, it is sufficient to keep $I_V \times N_V$ constant, so $\lambda$ will be constant in the system.

%-----------------------------------------------------------------------------------------------
% Acknowledgement
%-----------------------------------------------------------------------------------------------
\begin{acknowledgments}
PJZ would like to acknowledge the support of a project that has received funding from the European Union’s Horizon 2020 research and innovation programme under the Marie Sk{\l}odowska-Curie grant agreement No. 847413
and was a part of an international co-financed project founded from the programme of the Minister of Science and Higher Education entitled "PMW" in the years 2020 - 2024; agreement no. 5005/H2020-MSCA-COFUND/2019/2.
\end{acknowledgments}

\appendix

\section{Derivation of the Energy Expression} \label{appen1}
Here, we provide a derivation of the energy  of the system Eq.~(\ref{eqn:tot-e})  and of subsystems Eq.~(\ref{eqn:sub-e}). 

As stated in the main text, the energy density satisfies Eq.~(\ref{eqn:ss-state-eqn}). 
By moving the temperature profile to the left hand side (as $T(\vec{r}) > 0$) and integrate over the whole volume, the dependence over the particle density profile $n(\vec{r})$ is eliminated, 
\begin{equation}
\epsilon\int_{V}\ddfrac{\mathd^{3} \vec{r}}{T(\vec{r})} = \dfrac{3}{2}k_{B} \int_{V} \mathd^{3} \vec{r} n(\vec{r}) = \dfrac{3}{2}Nk_{B}.   
\label{eqn:app-1}
\end{equation}
From this relation, an expression of the energy density can be obtained, 
\begin{equation}
\epsilon = \dfrac{3}{2}N k_{B} \ddfrac{1}{\int_{V} \ddfrac{\mathd^{3}\vec{r}}{T(\vec{r})}}.  
\label{eqn:app-e}
\end{equation}
 Analogously, the energy density of the subsystem is 
\begin{equation}
\epsilon_{i} = \dfrac{3}{2}N_{i} k_{B} \ddfrac{1}{\int_{V_{i}} \ddfrac{\mathd^{3}\vec{r}}{T_{i}(\vec{r})}}.  
\label{eqn:app-sub-e}
\end{equation}

The temperature profile is obtained from Eq.~(\ref{eqn:laplace}) with the appropriate boundary conditions.   
Since the movable wall model is assumed to be infinite in $y$ and $z$ directions, it is sufficient to consider the dependence in $x$ direction, so  one has 
\begin{equation}
- k \dfrac{\partial^{2}}{\partial x^{2}} T(x) = \lambda. 
\end{equation}
The boundary conditions prior to the constraint are $T(\pm L) = T_{0}$, giving
\begin{equation}
T(x) = - \dfrac{\lambda}{2k} x^{2} + \dfrac{\lambda}{2k}L^{2} + T_{0}, 
\label{eqn:app-T}
\end{equation}
The additional boundary conditions under the constraint is $\partial_{x}T_{i}(x_{i}) = 0$, giving 
\begin{equation}
T_{i}(x) = - \dfrac{\lambda}{2k} (x - x_w)^{2} + \dfrac{\lambda}{2k}(L - x_w)^{2} + T_{0}. 
\label{eqn:app-Ti}
\end{equation}

Inserting Eq.~(\ref{eqn:app-T}) into Eq.~(\ref{eqn:app-e}), and Eq.~(\ref{eqn:app-Ti}) into Eq.~(\ref{eqn:app-sub-e}), the final expressions of energy are obtained, 
\begin{equation}
U = U_{eq} f(\lambda \cdot \ddfrac{L^{2}}{kT}) = \dfrac{3}{2}Nk_{B}T f(\lambda \cdot \dfrac{L^{2}}{kT}), 
\label{eqn:app-u}
\end{equation}
\begin{equation}
U_{i} = U_{i,eq} f(\lambda \cdot \dfrac{L_{i}^{2}}{kT}) = \dfrac{3}{2}N_{i}k_{B}T f(\lambda \cdot \dfrac{L_{i}^{2}}{kT}). 
\label{eqn:app-ui}
\end{equation}
where $L_{1} = L + x_w$, $L_{2} = L - x_w$ and $f(x) \equiv \sqrt{x(x+2)}/(2\arctanh \sqrt{x/(x+2)})$, as  in Eqs.~(\ref{eqn:tot-e})  and  Eq.~(\ref{eqn:sub-e}). 

\section{Derivation of the phase transition point $\lambda_{c}L^{2}/kT$}\label{app:critical-point}
%-----------------------------------------------------------------------------------------------
Here we provide a derivation of the transition point $\lambda_{c}L^{2}/kT$ of the movable wall model with $N_{1} = N_{2} = N/2$, where it is stated that  $\lambda_{c}L^{2}/kT \approx 4.55344$. 

We start by rewriting Eq.~(\ref{eqn:app-ui}) using the normalised variables $\til{\lambda} \equiv \lambda L^{2}/kT$ and $x_{w} = x_w/L$, and let $N_{i} = N/2$. Next, a new function is defined as the negative difference between energy densities, 
\begin{align}
&G(x_{w}) \equiv - (\epsilon_{1} - \epsilon_{2}) \equiv -\dfrac{3N k_{B}T}{2V} ( g(x_{w}) - g(-x_{w}) ), \label{eqn:app-G} \\
&g(x_{w}) = \ddfrac{f\left(\til{\lambda}(1 + x_{w})^2\right)}{1+x_{w}}.   
\end{align}
The negativity of $G(x_{w})$ is not necessary, but it is chosen here so that it would simplify the explanation in a later section. Since the function is odd with respect to $x_{w}$, it is sufficient to look at half of the axis, say $x_{w} \in [0, 1)$.

Consider the range $x_{w} \geq 0$. For this movable wall model with equal subsystem particles, the phase transition occurs when the number of solutions transit from $1$ to $2$. Equivalently, this means that the number of times $\epsilon_{1}(x_w)$ crosses with $\epsilon_{2}(x_w)$ in $x_w \geq 0$ transit from $1$ to $2$, which is then equivalently the crossings of $G(x)$ with the $x$-axis. 

More precisely, in the range $x_{w} \geq [0, 1)$, $G(0) = 0$ is fixed and $\lim_{x \rightarrow 1}G(x) \rightarrow \infty$. Therefore, when $G'(0) > 0$, $G(x)$ is monotonic and have only one crossing point at $x=0$; when $G'(0) < 0$, $G(x)$ will have $2$ crossing points; the transition point is $G'(0) = 0 = -2g'(0)$. 
Explicitly,  
\begin{equation}
\dfrac{\mathd g}{\mathd x}\bigg|_{x=0} = \ddfrac{1}{2+\til{\lambda}} - \ddfrac{\til{\lambda}^{2}\arctanh(\sqrt{\til{\lambda}/(\til{\lambda}+2)})}{(\til{\lambda}(\til{\lambda}+2))^{3/2}} = 0. 
\end{equation}
Solving this implicit equation numerically, the solution is $\til{\lambda}_{c} = \lambda_{c}L^{2}/kT \approx 4.55344$. 

%-----------------------------------------------------------------------------------------------
\section{Molecular dynamics simulation}\label{app:simulation}
%-----------------------------------------------------------------------------------------------
The simulations are performed using molecular dynamics method \cite{S1-allen1989computer} for  systems of $N = 153600$ particles of mass $m = 1$ enclosed in the rectangular box and interacting via the following potential:
\begin{equation}
V_{rep}(r) = \varepsilon\left(\dfrac{\sigma}{r}\right)^{12}
\end{equation}
where $r$ is the interparticle distance and both the energy, $\varepsilon$, and the size, $\sigma$, parameter are  set to one. The equations of motion are solved applying the Verlet method \cite{S1-allen1989computer} for the time step  $\delta t = 0.0025 \sigma(m/\varepsilon)^{1/2}$. The gas of particles is enclosed in the rectangular box of the edges: $L_{z} = L_{y} = 275.8, L_{x} \equiv 2L = 658.3$. The periodic boundary conditions were applied only along $z$ and $y$ axis. The $x$-th direction was restricted by two walls that repulse the particles with the potentials: $V_{rep}(-L-x)$ and $V_{rep}(L - x)$, where $-L \le x \le L$. The movable wall of the mass $M=32m$ is perpendicular to $x$-th axis and interacts with the surrounding particles with the potential: 
\begin{equation}
V_{w}(x) = 
\begin{cases}
V_{rep}(x_w + 1- x), &\text{for } x \leq x_w - 1\\
V_{rep}(x - x_w - 1), &\text{for } x \geq x_w + 1\\
\infty, &\text{for } x_{w} - 1 \leq x \leq x_w + 1
\end{cases}
\end{equation}
where $x_w$ denotes the position of the wall. At the beginning of each simulation run $x_w = 0 $ and the particles are equally distributed between the two parts.

\begin{center}
\begin{figure}[t!]
\includegraphics[width=0.4\textwidth]{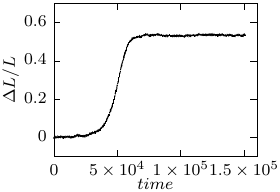}
%\vspace{-1cm}
\caption{The relative shift of the wall $\Delta L/L = \vert \langle x_{w} \rangle/L\vert$ as a function of time for  the mean rate of energy added per particle $\lambda_p=0.01$.}
\label{fig:S1}
\end{figure}
\end{center}

Energy is added to the system once per time interval $0.1\sigma(m/\varepsilon)^{1/2}$ and removed from the system by keeping the boundary temperature $T_{0}$ constant ($=0.5$) by applying Brownian simulations \cite{S1-allen1989computer}. For this purpose, the Verlet scheme is completed with the stochastic term \cite{S2-litniewski2006influence} for $x < -L + x_{T}$ and $x > L - x_{T}$ where $x_{T} = 3.0$. The system for $-L+x_{T} < x < L - x_{T}$ is imaginary divided into $20$ equal size layers perpendicular to the $x$-th axis. The energy flux is proportional to the  density \cite{Robert} \textit{i.e.}, the same amount of energy is added to the same volume (here, the layer) and equally shared between all particles in the layer.  As the initial state for all simulation runs we adopted the system at the equilibrium state at the temperature $T=T_{0} = 0.5$. 
The simulations are run for minimum 100 000$\delta t$ to assure that the steady state is achieved - see Fig.~\ref{fig:S1}.

%-----------------------------------------------------------------------------------------------
\section{Derivation of the steady-state functions for the movable wall model}\label{app:consist-check}
%-----------------------------------------------------------------------------------------------

Using the movable wall model, in particular Eq.~(\ref{eqn:tot-e}) and  Eq.~(\ref{eqn:sub-e}), we provide the derivation of the expressions for  $B$, $S, P$ and $\mu$ (Eqs.~(\ref{eqn:B})-(\ref{eqn:ss-mu}) , respectively).
This is done by using  consistency   relations, which are analogs of Maxwell relations in equilibrium thermodynamics. Further, we derive the integration form of $B$ (Eq.~(\ref{eqn:b-int})). 

In analogy to the Maxwell relations of equilibrium thermodynamics, in order for the proposed $3$ postulates concerning state function $B$  (described in the main text) 
to be valid, the following $6$ relations of mixed derivatives must be satisfied, 
\begin{gather}
\dfrac{\partial^{2} B}{\partial T \partial \lambda} = \dfrac{\partial^{2} B}{\partial \lambda \partial T} \Leftrightarrow 
\dfrac{\partial S}{\partial \lambda} = \dfrac{\partial X}{\partial T}, \label{eqn:app-sx}\\
\dfrac{\partial^{2} B}{\partial V \partial \lambda} = \dfrac{\partial^{2} B}{\partial \lambda \partial V} \Leftrightarrow 
\dfrac{\partial P}{\partial \lambda} = \dfrac{\partial X}{\partial V}, \label{eqn:app-px}\\
\dfrac{\partial^{2} B}{\partial N \partial \lambda} = \dfrac{\partial^{2} B}{\partial \lambda \partial N} \Leftrightarrow  
- \dfrac{\partial \mu}{\partial \lambda} = \dfrac{\partial X}{\partial N}, \label{eqn:app-mux} \\
\dfrac{\partial^{2} B}{\partial T \partial V} = \dfrac{\partial^{2} B}{\partial V \partial T} \Leftrightarrow 
\dfrac{\partial S}{\partial V} = \dfrac{\partial P}{\partial T},  \label{eqn:app-sp}\\
\dfrac{\partial^{2} B}{\partial T \partial N} = \dfrac{\partial^{2} B}{\partial N \partial T} \Leftrightarrow  
- \dfrac{\partial S}{\partial N} = \dfrac{\partial \mu}{\partial T}, \label{eqn:app-smu}\\
\dfrac{\partial^{2} B}{\partial V \partial N} = \dfrac{\partial^{2} B}{\partial N \partial V} \Leftrightarrow 
- \dfrac{\partial P}{\partial N} = \dfrac{\partial \mu}{\partial V}. \label{eqn:app-pmu} 
\end{gather}

From Eq.~(\ref{eqn:app-sx}), the steady state expression of entropy $S$ can be obtained from 
%\begin{equation}
%\dfrac{\partial X}{\partial T} = \ddfrac{N k_{B}}{2}\ddfrac{f(\lambda \ddfrac{L^{2}}{kT}) -1}{\lambda}, 
%\end{equation}
%and integrate with respect to $\lambda$, we obtain eqn.(\ref{eqn:ss-s}), 
\begin{equation}
\begin{aligned}
S(T, V, N, \lambda) =& \int_{0}^{\lambda} \dfrac{\partial X}{\partial T} d\lambda' + S_{eq}(T, V, N) \\
=&\ddfrac{N k_{B}}{2}\int_{0}^{\lambda} \ddfrac{f(\lambda \ddfrac{L^{2}}{kT}) - 1}{\lambda} \mathd \lambda \\
&  - \ddfrac{N k_{B}}{2} (f (\lambda \ddfrac{L^{2}}{kT}) - 1) + S_{eq} (T, V, N), \\
\end{aligned}
\label{eqn:app-sss}
\end{equation}
which is Eq,~(\ref{eqn:ss-s}). 
Further, from postulate Eq.~(\ref{eqn:B-diff})  that $\partial B/\partial T \equiv - S$, $B$ is given by
\begin{equation}
\begin{aligned}
&B(T, V, N, \lambda) - B(T_{ref}, V, N, \lambda) = - \int_{T_{ref}}^{T} SdT'  \\
&= - \int_{T_{ref}}^{T} dT' \left(\int_{0}^{\lambda} \ddfrac{\partial X}{\partial T'}d\lambda' + S_{eq}(T', V, N)\right), 
\end{aligned}
\end{equation}
Changing the order of integration, the above expression becomes 
\begin{equation}
\begin{aligned}
&B(T, V, N, \lambda) - B(T_{ref}, V, N, \lambda) \\  &= F_{eq}(T, V, N) - F_{eq}(T_{ref}, V, N)  \\
&- \int_{0}^{\lambda} X(T, V, N, \lambda') d\lambda' + \int_{0}^{\lambda} X(T_{ref}, V, N, \lambda') d\lambda', 
\end{aligned}
\end{equation}
Thus, we conclude:

 \begin{equation}
B(T, V, N, \lambda) = F_{eq}(T, V, N) - \int_{0}^{\lambda} X d\lambda' . 
\label{eqn:app-B}
\end{equation}

Next, from Eq.~(\ref{eqn:app-px}), we obtain $P$ 
\begin{equation}
P = \int_{0}^{\lambda} \ddfrac{\partial X}{\partial V} d\lambda' + P_{eq}(T, V, N) = \dfrac{Nk_{B}T}{V} f(\lambda \dfrac{L^{2}}{kT}). 
\label{eqn:app-pss}
\end{equation}
as given by Eq.~(\ref{eqn:ss-p}). 
This expression is consistent with the ideal gas law where $P= 3U/2V$, and with the definition from $P \equiv -\partial B/\partial V$ where $B$ is given by Eq.~(\ref{eqn:app-B}).

Thirdly, from relation Eq.~(\ref{eqn:app-mux}), we obtain $\mu$
\begin{equation}
\begin{aligned}
\mu(T, V, N, \lambda) &= - \int_{0}^{\lambda}\dfrac{\partial X}{\partial N}d\lambda' + \mu_{eq}(T, V, N), \\
&= \ddfrac{k_{B}T}{2}\int_{0}^{\lambda} \ddfrac{f(\lambda \dfrac{L^{2}}{kT}) - 1}{\lambda} \mathd \lambda + \mu_{eq}(T, V, N)
\end{aligned}
\label{eqn:app-muss}
\end{equation}
as shown in Eq.~(\ref{eqn:ss-mu}). Similarly, this expression is consistent with the definition $\mu \equiv \partial B/\partial N$. 

Now, we consider the  rest of the relations, i.e.,  Eqs.~(\ref{eqn:app-sp}, \ref{eqn:app-smu}, \ref{eqn:app-pmu}). Using the above expressions of the state functions, and obtain
\begin{gather}
%sp
\dfrac{\partial S}{\partial V} = \dfrac{\partial P}{\partial T} = \dfrac{Nk_{B}f}{V} - \dfrac{Nk_{B}}{V}\dfrac{\lambda L^{2}}{kT}\dfrac{d f}{d y},  \\
%smu
- \dfrac{\partial S}{\partial N} = \dfrac{\partial \mu}{\partial T} = -\dfrac{k_{B}}{2}\int_{0}^{\lambda}\dfrac{f-1}{\lambda'}d\lambda' + \dfrac{k_{B}}{2}(f-1) + \dfrac{\partial \mu_{eq}}{\partial T},  \\
%pmu
- \dfrac{\partial P}{\partial N} = \dfrac{\partial \mu}{\partial V} = - \dfrac{k_{B}T}{V}f, 
\end{gather}
where $y = \lambda L^{2}/kT$, and we have used the equilibrium relation $\partial S_{eq}/\partial N = -\partial \mu_{eq}/\partial T$. %from the Maxwell relations of the Gibbs free energy.  

Finally, using postulate Eq.~(\ref{eqn:x}) and the above obtained Eqs.~(\ref{eqn:app-sss}, \ref{eqn:app-pss}, \ref{eqn:app-muss}), the integration form of $B$ can be written as 
\begin{equation}
B = U - TS - 4X\lambda,  
\end{equation}
which is Eq.~(\ref{eqn:b-int}), in order to be consistent with Eq.~(\ref{eqn:app-B}).  

%-----------------------------------------------------------------------------------------------
\section{Analysis of the extrema of $B$}\label{app:extremum}
%-----------------------------------------------------------------------------------------------
Here we check the properties of the extrema of $B$ of the movable wall model with $N_{1} = N_{2} = N/2$. Equivalently, it is to check the second order derivative $\partial^{2} B/\partial x_w^{2}$ at extrema $x_w^{*}$, which are solutions to $\partial B/\partial x_w = 0$. In other words, $x_w^{*}$ are local maxima if $(\partial^{2} B/\partial x_w^{2})(x_w^{*}) < 0$, and local minima if $(\partial^{2} B/\partial x_w^{2}) (x_w^{*}) > 0$. 

Using $G(x_{w})$ in Eq.~(\ref{eqn:app-G}), the comparison between the second derivative of $B$ and $0$ simplifies to comparison between
$ - \dfrac{\partial }{\partial x_{w}} \left( \ddfrac{f (\til{\lambda}(1 + x_{w})^2)}{1+x_{w}} - \ddfrac{f(\til{\lambda} (1 + x_{w})^2)}{1-x_{w}}\right)\bigg|_{x_{w}^{*}}$ and  $0$ or, equivalently, between
$G'(x_{w})\bigg|_{x_{w}^{*}}$  and $0$.  
As shown earlier,  beyond the transition point we have $G'(0) <0$, suggesting that $x_w^{*} = 0$ corresponds to a local maximum. Moreover, since $G(1) > G(0)$, the derivative at the crossing point $x_w^{*} > 0$ must be positive, $G(x_{w}^{*}>0) > 0$, suggesting that this solution is a local minimum. Finally, since $G(x)$ is an odd function,  $G'(-x_{w}) = G'(x_{w})$ and the crossing point $x_w^{*} < 0$ is also a local minimum. The above discussion shows that $x_w^{*} = 0$ is the local maximum and $x_w^{*} \neq 0$ are local minima. 

%-----------------------------------------------------------------------------------------------
% Bibliography
%-----------------------------------------------------------------------------------------------
\bibliographystyle{bibgen}
\bibliography{bibliography}

\begin{thebibliography}{10}
\newcommand{\enquote}[1]{``#1''}

\bibitem{groot1962}
S.~R. de~Groot and P.~Mazur, {\em Non-Equilibrium Thermodynamics\/},
  North-Holland, Amsterdam (1962).

\bibitem{prigogine}
D.~Kondepudi and I.~Prigogine, {\em Modern thermodynamics: from heat engines to
  dissipative structures\/}, Wiley, New York, second edition (1998).

\bibitem{oono1998steady}
Y.~Oono and M.~Paniconi, \enquote{Steady state thermodynamics}, {\em Progress
  of Theoretical Physics Supplement\/} {\bf 130}, 29–44 (1998).

\bibitem{Sasa:2006}
S.~Sasa and H.~Tasaki, \enquote{Steady state thermodynamics.}, {\em J. Stat.
  Phys.\/} {\bf 125}, 125 (2006).

\bibitem{revUdo}
U.~Seifert, \enquote{Stochastic thermodynamics, fluctuation theorems and
  molecular machines}, {\em Rep. Prog. Phys.\/} {\bf 75}, 126001 (2012).

\bibitem{holyst2012thermodynamics}
R.~Ho{\l}yst and A.~Poniewierski, {\em Thermodynamics for chemists, physicist
  and engineers\/}, Springer Science \& Business Media (2012).

\bibitem{onsager1}
L.~Onsager, \enquote{Reciprocal relations in irreversible processes. i.}, {\em
  Phys. Rev.\/} {\bf 37}, 405 (1931).

\bibitem{onsager2}
L.~Onsager, \enquote{Reciprocal relations in irreversible processes. ii.}, {\em
  Phys. Rev.\/} {\bf 38}, 2265 (1931).

\bibitem{ziegler}
H.~Ziegler, {\em An introduction to thermomechanics\/}, Elsevier (2012).

\bibitem{PhysRevE.80.021113}
R.~K. Niven, \enquote{Steady state of a dissipative flow-controlled system and
  the maximum entropy production principle}, {\em Phys. Rev. E\/} {\bf 80},
  021113 (2009).

\bibitem{Dewar2014}
R.~C. Dewar and A.~Maritan, \enquote{A theoretical basis for maximum entropy
  production}, in \enquote{Beyond the Second Law: Entropy Production and
  Non-equilibrium Systems}, pp. 49--71, Springer Berlin Heidelberg, Berlin,
  Heidelberg (2014).

\bibitem{wang2006maximum}
Q.~Wang, \enquote{Maximum entropy change and least action principle for
  nonequilibrium systems}, {\em Astrophysics and Space Science\/} {\bf 305},
  273 (2006).

\bibitem{martyushev2013entropy}
L.~M. Martyushev, \enquote{Entropy and entropy production: Old misconceptions
  and new breakthroughs}, {\em Entropy\/} {\bf 15}, 1152–1170 (2013).

\bibitem{dewar2014beyond}
R.~C. Dewar, C.~H. Lineweaver, R.~K. Niven, and K.~Regenauer-Lieb,
  \enquote{Beyond the second law: An overview}, in \enquote{Beyond the Second
  Law: Entropy Production and Non-equilibrium Systems}, pp. 3--27, Springer,
  Berlin, Heidelberg (2014).

\bibitem{endres2017entropy}
R.~G. Endres, \enquote{Entropy production selects nonequilibrium states in
  multistable systems}, {\em Sci. Rep.\/} {\bf 7}, 1 (2017).

\bibitem{doi:10.1063/1.2400859}
P.~Attard, \enquote{Statistical mechanical theory for steady state systems. vi.
  variational principles}, {\em J. Chem. Phys.\/} {\bf 125}, 214502 (2006).

\bibitem{revUdo01}
U.~Seifert, \enquote{Entropy production along a stochastic trajetory and an
  integral fluctuation theorem}, {\em Phys. Rev. Lett.\/} {\bf 95}, 040602
  (2005).

\bibitem{gallavotti1995dynamical}
G.~Gallavotti and E.~G. Cohen, \enquote{Dynamical ensembles in nonequilibrium
  statistical mechanics}, {\em Phys. Rev. Lett.\/} {\bf 74}, 2694 (1995).

\bibitem{lebowitz1999gallavotti}
J.~L. Lebowitz and H.~Spohn, \enquote{A gallavotti--cohen-type symmetry in the
  large deviation functional for stochastic dynamics}, {\em J. Stat. Phys.\/}
  {\bf 95}, 333 (1999).

\bibitem{kurchan1998fluctuation}
J.~Kurchan, \enquote{Fluctuation theorem for stochastic dynamics}, {\em J.
  Phys. A: Math. Gen.\/} {\bf 31}, 3719 (1998).

\bibitem{evans1994equilibrium}
D.~J. Evans and D.~J. Searles, \enquote{Equilibrium microstates which generate
  second law violating steady states}, {\em Phys. Rev. E\/} {\bf 50}, 1645
  (1994).

\bibitem{jarzynski}
C.~Jarzynski, \enquote{Nonequilibrium equality for free energy differences},
  {\em Phys. Rev. Lett.\/} {\bf 78}, 2690 (1997).

\bibitem{crooks2000path}
G.~E. Crooks, \enquote{Path-ensemble averages in systems driven far from
  equilibrium}, {\em Phys. Rev. E\/} {\bf 61}, 2361 (2000).

\bibitem{Robert}
R.~Ho{\l}yst, A.~Macio{\l}ek, Y.~Zhang, M.~Litniewski, P.~Knycha{\l}a,
  M.~Kasprzak, and M.~Banaszak, \enquote{Flux and storage of energy in
  non-equilibrium,stationary states}, {\em Phys. Rev. E\/} {\bf 99}, 042118
  (2019).

\bibitem{zhang2020energy}
Y.~Zhang, R.~Ho{\l}yst, and A.~Macio{\l}ek, \enquote{Energy storage in steady
  states under cyclic local energy input}, {\em Phys. Rev. E\/} {\bf 101},
  012127 (2020).

\bibitem{zhang2020storage}
Y.~Zhang, K.~Gi\.zy\'nski, A.~Macio{\l}ek, and R.~Ho{\l}yst, \enquote{Storage
  of energy in constrained non-equilibrium systems}, {\em Entropy\/} {\bf 22},
  557 (2020).

\bibitem{Niven2010}
R.~K. Niven, \enquote{Minimization of a free-energy-like potential for
  non-equilibrium flow systems at steady state}, {\em Phil. Trans. R. Soc. B\/}
  {\bf 365}, 1323–1331 (2010).

\bibitem{komatsu2008steady}
T.~S. Komatsu, N.~Nakagawa, S.-i. Sasa, and H.~Tasaki, \enquote{Steady-state
  thermodynamics for heat conduction: microscopic derivation}, {\em Phys. Rev.
  Lett.\/} {\bf 100}, 230602 (2008).

\bibitem{S1-allen1989computer}
M.~Allen and D.~Tildesley, {\em Computer Simulation of Liquids\/}, Oxford
  Science Publications, Clarendon Press (1989).

\bibitem{S2-litniewski2006influence}
M.~Litniewski, \enquote{The influence of the quencher concentration on the rate
  of simple bimolecular reaction: Molecular dynamics study. \text{II}}, {\em J.
  Chem. Phys.\/} {\bf 124}, 114501 (2006).

\end{thebibliography}

\end{document}